# Nanophotonic cavity optomechanics with propagating phonons in microwave Ku band

Huan Li[1*], Semere A. Tadesse[1,2 *], Qiyu Liu[1], and Mo Li[1†]

[1]Department of Electrical and Computer Engineering, University of Minnesota, Minneapolis, MN 55455, USA
[2]School of Physics and Astronomy, University of Minnesota, Minneapolis, MN 55455, USA

**Sideband-resolved coupling between multiple photonic nanocavities and propagating mechanical waves in microwave Ku-band is demonstrated. Coherent and strong photon-phonon interaction is manifested with optomechanically induced transparency and absorption, and phase-coherent interaction in multiple cavities. Inside an echo chamber it is shown that a phonon pulse can interact with an embedded nanocavity for multiple times. Our device provides a scalable platform to optomechanically couple phonons and photons for microwave photonics and quantum photonics.**

Strong coherent interactions between co-localized optical and mechanical eigenmodes of various cavity optomechanical systems has been intensively explored toward quantum information processing using both photons and phonons[1-5]. To achieve this, optomechanically induced transparency[6-8], backaction cooling[9, 10] and light squeezing[11, 12] have been demonstrated. To this end, optomechanical interaction between localized mechanical modes and confined optical modes of the cavity system is exploited to obtain strong photon-phonon coupling. In contrast to localized motion, however, the more ubiquitous form of phonons is various types of mechanical waves that freely propagate in the bulk or on the surface of solids. Propagating mechanical waves can interact with light strongly through elasto-optic effects and have been utilized in various acousto-optic

---

devices[13, 14]. In optical fibers[15-18] and more recently in integrated waveguides[19-23], mechanical waves can also be optically stimulated and lead to strong Brillouin scattering between phase-matched optical modes. Much like photons, these mechanical waves, or itinerant phonons, can be confined and guided in planar phononic waveguides to propagate over long distances so that their coupling with photonic modes are highly scalable[24-26]. Furthermore, phononic crystals and cavities can also be implemented to confine and store phonons to achieve extended lifetime[27, 28]. The quantum nature of those itinerant mechanical states has recently been revealed[29].

While both propagating and localized mechanical modes can be optically stimulated, they can be more efficiently excited electromechanically so that the amplitude or number of phonons can be pumped to a very high level without the need of a strong pump laser. Such approach has been employed in electro-optomechanical systems that can directly couple microwave signals with optical signals[30, 31]. Mechanical waves can also be excited with electromechanical transducers in monolithic devices. Representative types of mechanical waves include bulk acoustic waves (BAW), Rayleigh surface acoustic waves (SAW), Lamb waves and flexural plate waves. Rayleigh SAW waves are particularly interesting because the displacement is confined on the surface of the substrate and it can be conveniently excited with planar transducers. Acoustic wave devices have long been applied for wireless communication and signal processing applications[32, 33]. In quantum physics, single quanta of SAW wave has been detected with single electron transistors and superconducting qubits, manifesting that propagating phonon can be a viable quantum information carrier[29, 34]. In photonics, we previously showed that microwave frequency SAW wave can efficiently modulate photonic waveguide modes[35]. In this work, we present a new type of cavity optomechanical system consisting of high-Q photonic crystal nanocavities integrated with SAW transducers working at frequency above 12 GHz, entering the microwave Ku-band. At this unprecedentedly high frequency, optomechanical interaction in the new system reaches the sideband resolved regime and enables optomechanically induced transparency and absorption. To establish the propagating feature of mechanical waves, we demonstrate phase-coherent optomechanical modulation of multiple nanocavities and that an acoustic pulse travels insides an echo chamber for multiple rounds to interact with an embedded nanocavity.

Both the photonic crystal nanocavities and the SAW transducers are integrated on 330 nm thick aluminum nitride (AlN) film, which provides strong piezoelectricity for efficient excitation of SAW waves and high refractive index for optical confinement. As illustrated in Fig. 1a, the nanocavity is formed in a nanobeam inscribed with one dimensional photonic crystal shallow etched in the AlN layer. The nanocavity can be coupled with a waveguide either on the side or at the ends. An interdigital transducer (IDT) is configured to launch surface acoustic wave propagating in the transverse direction to the nanobeam. The aperture of the IDT (100 μm) is designed to be much larger than the mode size of the nanocavity with 24 μm full-width at the half-maximum (FWHM) of the Gaussian intensity profile, as seen in Fig. 1b. Therefore the nanocavity undergoes approximately uniform deformation when the acoustic wave passes. The wavelength of the acoustic wave that is excited is determined by the period of the IDT transducer while the frequencies of the modes are strongly dependent on their dispersion properties in the multilayer substrate[23]. The scanning electron microscope image in Fig. 1c shows the IDT electrodes with period of 450 nm and linewidth of 112.5 nm. To achieve high optical quality factor and strong optomechanical interaction, we optimized the photonic crystal nanocavity design such that the electric field of the fundamental dielectric mode[36] is well confined inside the AlN structures with an effective mode index of $n_{\text{eff}} \approx 1.54$. (The design of the photonic crystal nanocavity is discussed in the Supplementary Information.) Fig. 1d displays measured transmission spectrum of the nanocavity side coupled with a waveguide, showing the fundamental and the first order cavity modes as dips. The quality factor of the waveguide loaded fundamental mode is $5\times10^4$, corresponding to a linewidth of $\kappa=(2\pi)\cdot 3.88$ GHz (Fig. 1e). To characterize the SAW modes, reflection coefficient ($S_{11}$) of the IDT was measured with a vector network analyzer (VNA) and plotted in Fig. 1f. When an acoustic wave is excited, the reflection spectrum also shows a negative peak as the microwave signal is converted to outgoing waves so less reflected. Prominent resonance peaks corresponding to high order Rayleigh modes, as marked in Fig. 1f, can be observed in the reflection spectrum. These high order modes are more interesting than the low order modes because mechanical energy is more confined in the AlN layer rather than the $SiO_2$ so that their overlap with the cavity optical mode is more significant, inducing stronger optomechanical coupling[23]. The inset of Fig. 1f shows simulated mode profile of Rayleigh mode R14 with more than 20% mechanical energy confined in the AlN layer. Its frequency is $\Omega_{14}=(2\pi)\cdot 12.1$ GHz, entering the microwave Ku band. The linewidth of the mode resonance is $\Gamma_{14}=(2\pi)\cdot 38.9$ MHz (Fig. 1g). We note that one feature of our platform is that the electromechanical transducer is separated from the photonic cavity, allowing the mechanical frequency to be freely engineered and the photonic cavity to be independently optimized. (More simulation results of the SAW modes are included in the Supplementary Information.)

To characterize the optomechanical coupling between the nanocavity and the SAW waves, a laser source, with a variable detuning from the fundamental cavity mode, was sent into the input waveguide with 22 µW power. The transmitted optical power was measured with a high-speed photodetector which was connected to the VNA and the overall system transmission coefficient $S_{21}$ was measured as a function of the excitation frequency at the IDT. Broadband optical $S_{21}$ spectrum is shown in Fig. 2a. It can be seen that in addition to the Rayleigh modes (R11–R16) observable in $S_{11}$, optical $S_{21}$ also detects additional modes (R4–R10) that are not visible in $S_{11}$ spectrum. Comparison of the two spectra demonstrates that the high-Q nanocavity provides optical detection of acoustic waves with broader bandwidth and higher sensitivity. The amplitude of the $S_{21}$ peaks are proportional to the modal overlaps of the Rayleigh modes with the cavity mode[23]. Fig. 2b shows zoom-in of the $S_{11}$ and $S_{21}$ of mode R14. Since the frequency of the mechanical mode ($\Omega_{12}=(2\pi)\cdot 12.1$ GHz) is considerably greater than the dissipation rate of the nanocavity mode $\kappa=(2\pi)\cdot 3.88$ GHz, their optomechanical coupling is in the sideband resolved regime, a prerequisite condition for phenomena such as induced transparency and strong coupling. Fig. 2c shows the $S_{21}$ peak amplitude of mode R14 as a function of varying laser-cavity detuning, displaying the characteristic lineshape of sideband resolved cavity optomechanical coupling. Fitting the results with theoretical model and calibrating the transducing factors of the system provide the optomechanical coupling coefficient of the system, $G=(2\pi)\cdot 53$ GHz/nm, or expressed in the power of SAW, $(2\pi)\cdot 23$ MHz/µW$^{1/2}$. (More details are provided in the Supplementary Information).

Coherent interaction between cavity photons and propagating phonons generates Stokes and anti-Stokes photons, which can interfere with probe photons constructively (destructively) to induce optical transparency (absorption). This three-wave nonlinear process is illustrated in the diagram of Fig. 3a. Different from optically stimulated phonons in conventional cavity optomechanics and stimulated Brillouin scattering (SBS), here the phonons are electromechanically excited non-locally and propagating. We investigate coherent photon-phonon interaction in our system using the setup depicted in Fig. 3b. Briefly, a laser is detuned from the cavity resonance by exactly the SAW mode frequency ($\Omega_{SAW}$) to provide the control light at frequency $\omega_c$, and modulated with an electro-optic modulator to generate sidebands with the upper one at frequency $\omega_p = \omega_c + \Delta_p$ used as the probe light. With this scheme, the transmission spectrum of the cavity can be measured by varying the modulation frequency to scan $\omega_p$ and detecting the beating signal between the transmitted probe light and the control light. The result is displayed as the grey symbols in Fig. 3c, showing a transmission dip which can be understood as cavity absorption. When the modulation signal is also sent to drive the IDT, a SAW wave of the same frequency is excited and propagates to the nanocavity to couple with the control light. This optomechanical coupling leads to three-wave mixing between the control, the probe and the SAW wave. Depending on the SAW phase which can be controlled with a phase shifter, the interference of waves leads to transparency or absorption. When the interference is constructive (destructive), a transparency (absorption) window is observed within the cavity resonance, seen as the red peak (blue dip) in Fig. 3c. Because in this homodyne measurement scheme the mechanical frequency and the probe detuning are synchronized, the transparency (absorption) window width agrees with the SAW IDT bandwidth (Fig. 2b). Fixing the control light power, the SAW wave can be excited to a high amplitude to compensate the cavity loss and even amplify the probe light with a considerable gain, as shown in Fig. 3d. When the gain is high so the cavity absorption is negligible, it is proportional to the SAW power (or number of propagating phonons), as shown in Fig. 3f. In addition to transparency or absorption, the three-wave mixing process is controlled by the phase of the SAW wave relative to the probe. Columns in Fig. 3e show the situations when the phase shift $\phi$ is set to 0, $\pi/2$, $\pi$ and $3\pi/2$, respectively, so that the interference is tuned from constructive to destructive and displays Fano-resonance-like lineshape in between.

      Besides the phenomena observable in sideband resolved cavity optomechanics, an important feature of our new platform is that the propagating mechanical wave can interact with multiple cavities in a coherent fashion. This scalability will be important, for example, to wavelength multiplexed coupling and conversion between microwave and optical photons. We demonstrate scalability by placing three nanocavities in the path of the SAW wave as shown in Fig. 4a. These cavities are end-coupled with the waveguides and SAW transducers operate at a lower frequency of 1.75 GHz. As the beam of SAW wave propagates, it also undergoes diffraction that can be described by integrating Lamb's point source solution along all the IDT finger pairs[37],

each of which is treated as an effective line source. Overlaid in Fig. 4a is the calculated displacement field amplitude of the propagating SAW wave, showing its diffraction pattern. Counter-intuitively, as shown in Fig. 4b, the displacement amplitude of the SAW wave changes non-monotonically along the central line of the IDT where the nanocavities lie. This is confirmed in the optical $S_{21}$ spectra measured from the three cavities, as shown in Fig. 4c, in which the farthest cavity (1.5 mm from the IDT) shows almost equally strong modulation as the second cavity (0.5 mm from the IDT). On the other hand, the phases of the optomechanical modulation of the three cavities are coherent with incremental time delays due to the propagating of the SAW wave. The phase delay and the distance of the three cavities to the IDT transducers are plotted in Fig. 4d. The slope in the plot indicates the group velocity of the SAW wave to be 4.0 km/s. The demonstrated coupling between multiple cavities and SAW wave with well understood diffraction can be utilized to implement multiplexed microwave signal processing in optical domain.

Finally, propagating phonons can be guided and confined in a fashion much like photons, with phononic structures such as one or two dimensional phononic crystals. Here we use acoustic Bragg reflectors to build a planar phononic cavity, or acoustic echo chamber, inside which a photonic nanocavity is inserted to investigate photon-phonon interaction. Optical image of the device is displayed in Fig. 5a. Fig. 5b shows the optical $S_{21}$ spectra measured with the nanocavity inside phononic cavities of different lengths $D$=0.3, 0.6 and 0.9 mm, respectively. The spectra show additional peaks within the IDT bandwidth, corresponding to the acoustic resonances of the phononic cavity. Similar to an optical Fabry-Perot cavity, the spacing between the peaks, or the free-spectral range, $\Delta\Omega$, decreases with increasing cavity length as given by $\Delta\Omega=2\pi/T=\pi c/(D+d)$, where $T$ is the round-trip time, $c$ is the group velocity of the SAW wave and $d$ is the effective extra cavity length due to the Bragg reflectors. The nanocavity provides highly sensitive and broadband detection of acoustic wave travelling inside the chamber. By performing time domain measurement, we show that the nanocavity can "hear" multiple echoes of an acoustic pulse bouncing inside the chamber, as displayed in Fig. 5c. The acoustic pulse is first excited by a 40 ns microwave burst sent to the IDT (orange). Due to electrical and optical delay, after ~160ns, the pulse was detected by the nanocavity for the first time. The pulse then propagates back and forth between the reflector and the IDT. It passes and is detected by the nanocavity for four times, as most obvious in the top panel, before its amplitude decays below the noise floor. From the arrival time and signal amplitude of multiple echoes, shown in Fig. 5d and e, we observe that, other than linear propagation delay and loss, extra delay of 100 ns (corresponding to $d$=200 μm) and loss of 8 dB occur during each round trip. Those are attributed to the Bragg reflector of finite length and can be optimized with more advanced design of low-loss phononic reflectors.

In conclusion, we demonstrate a planar cavity optomechanical platform on which propagating acoustic wave can be generated, confined and guided to interact with photonic cavities integrated in the same layer of aluminum nitride. By using high resolution electron beam lithography, IDT can be patterned to excite acoustic wave at frequency into the microwave Ku band. We expect it to be straightforward to further increase the frequency by using more advanced nanofabrication technique such as nanoimprint lithography. In addition to high frequency, an important feature of propagating mechanical wave is its scalability – multiple photonic cavities can be coupled. Also, for phononic cavities or echo chambers, high cavity finesse can be obtained if the acoustic loss can be reduced by removing the substrate leakage, reducing diffraction with acoustic waveguide[28, 38] and suppressing intrinsic material loss at cryogenic temperature[39]. Therefore, it is promising to achieve photon-phonon interaction in the regime of strong coupling on this new platform as a scalable modality of quantum optomechanics.

**Methods**

**Device Fabrication** The devices were fabricated from a c-axis oriented, 330 nm thick piezoelectric aluminum nitride thin film sputtered (OEM Group, AZ) on a silicon wafer with a 3 μm buried silicon dioxide layer. The photonics layer was first patterned by electron beam lithography (Vistec EBPG-5000+) using ZEP-520A resist followed by chlorine based reactive ion etching. The AlN layer was etched down by 200 nm, leaving 130 nm thick AlN slab for the SAW wave to propagate without significant reflection and loss. The SAW IDT electrodes and contact pads were fabricated by electron beam lithography, followed by a 35 nm Ti/Au deposition and a liftoff process.

**Measurement Setup** An external cavity tunable semiconductor laser was used as the laser source with its output power stabilized using a feedback loop. A 20 GHz electro-optic power modulator (EOM) was used to generate the probe sidebands for the observation of optomechanically induced transparency and absorption. The laser (and the probe sidebands, if present) was further conditioned with a fiber polarization controller (FPC) and a variable optical attenuator (VOA) before coupled into and out of the photonic crystal nanocavity via the on-chip grating couplers and waveguides. The output laser from the nanocavity was amplified with an erbium doped fiber amplifier (EDFA), filtered with an optical tunable band-pass filter (OTF) and measured with a high-speed photodetector (PD, New Focus 1474-A).

In frequency domain measurement, a vector network analyzer (VNA) (Agilent E8362B) was used to measure frequency response of the system. For $S_{11}$ measurement, the VNA Port 1 was directly connected to the on-chip IDT through a microwave probe. For $S_{21}$ measurements, the RF power output from VNA Port 1 was split into two paths with an RF power splitter. One path was connected to the on-chip IDT through the RF probe. The other path was connected to the EOM to generate the optical probe sidebands. Both paths were properly conditioned with RF amplifiers, tunable attenuators and/or tunable delay lines as phase shifters. The output from the PD was amplified before being sent into the VNA Port 2. In time domain measurements, RF bursts were generated by gating a continuous wave (CW) RF source with a pulse generator and a high-speed RF switch. The RF source and the pulse generator were properly synchronized to minimize phase jitter, which was crucial to ensure excellent identicalness of all the RF bursts. The generated RF bursts were sent to the on-chip IDT through the RF probe. The output of the PD was amplified and measured with an oscilloscope which was synchronously triggered by the pulse generator. To achieve high signal to noise ratio, the device responses shown in Fig. 5 were averaged for 2 seconds.

**Acknowledgements**

We acknowledge the funding support provided by the Young Investigator Program (YIP) of AFOSR (Award Number FA9550-12-1-0338) and the National Science Foundation (Award Number ECCS-1307601). Parts of this work was carried out in the University of Minnesota Nanofabrication Center which receives partial support from NSF through NNIN program, and the Characterization Facility which is a member of the NSF-funded Materials Research Facilities Network via the MRSEC program.

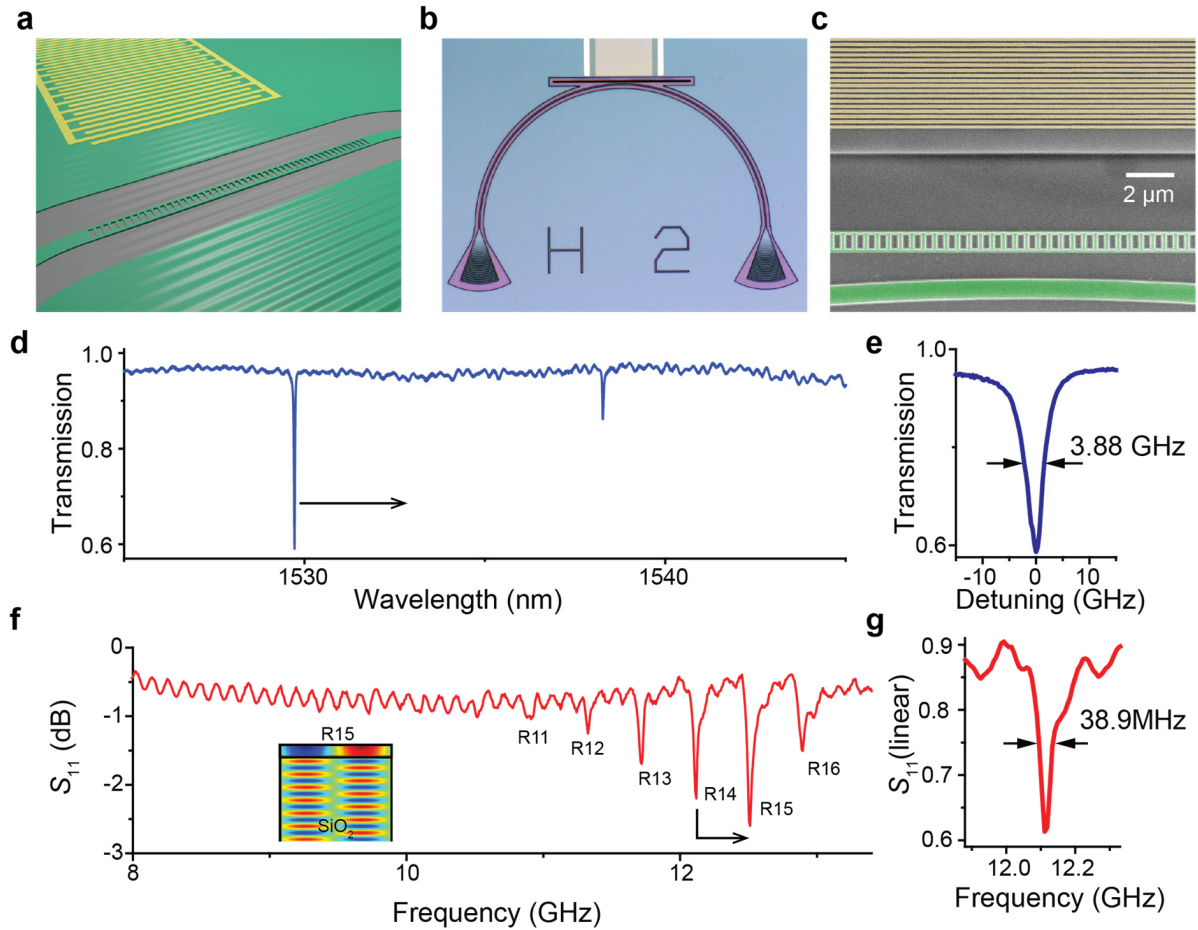

**Figure 1 Planarly integrated photonic crystal nanocavity and microwave frequency SAW wave transducer on AlN film. a)** 3D illustration of the device configuration, featuring the interdigital transducer and the excited SAW wave propagating in the transverse direction to the nanobeam photonic crystal nanocavity. **b)** Optical microscope image of a device. The nanocavity is side-coupled to a waveguide connected with two grating couplers. **c)** Scanning electron microscope image of the nanobeam cavity (green) and the IDT (yellow). The linewidth of the IDT figures is 112.5 nm. **d)** Transmission spectrum measured from the nanocavity showing the fundamental (at 1529.7 nm) and the first order (at 1538.3 nm) resonance modes. **e)** Zoom-in of the fundamental resonance of the nanocavity, showing a linewidth of 3.88 GHz. **f)** Spectrum of de-embedded and normalized reflection coefficient $S_{11}$ of the SAW IDT. High order Rayleigh modes from R11 to R16 can be observed as resonance dips. Inset: Simulated displacement field of R14 mode, showing the displacement is more confined in the top AlN layer. **g)** Zoom-in of the normalized $S_{11}$ spectrum of R14 mode plotted in linear scale, showing a linewidth of 38.9 MHz.

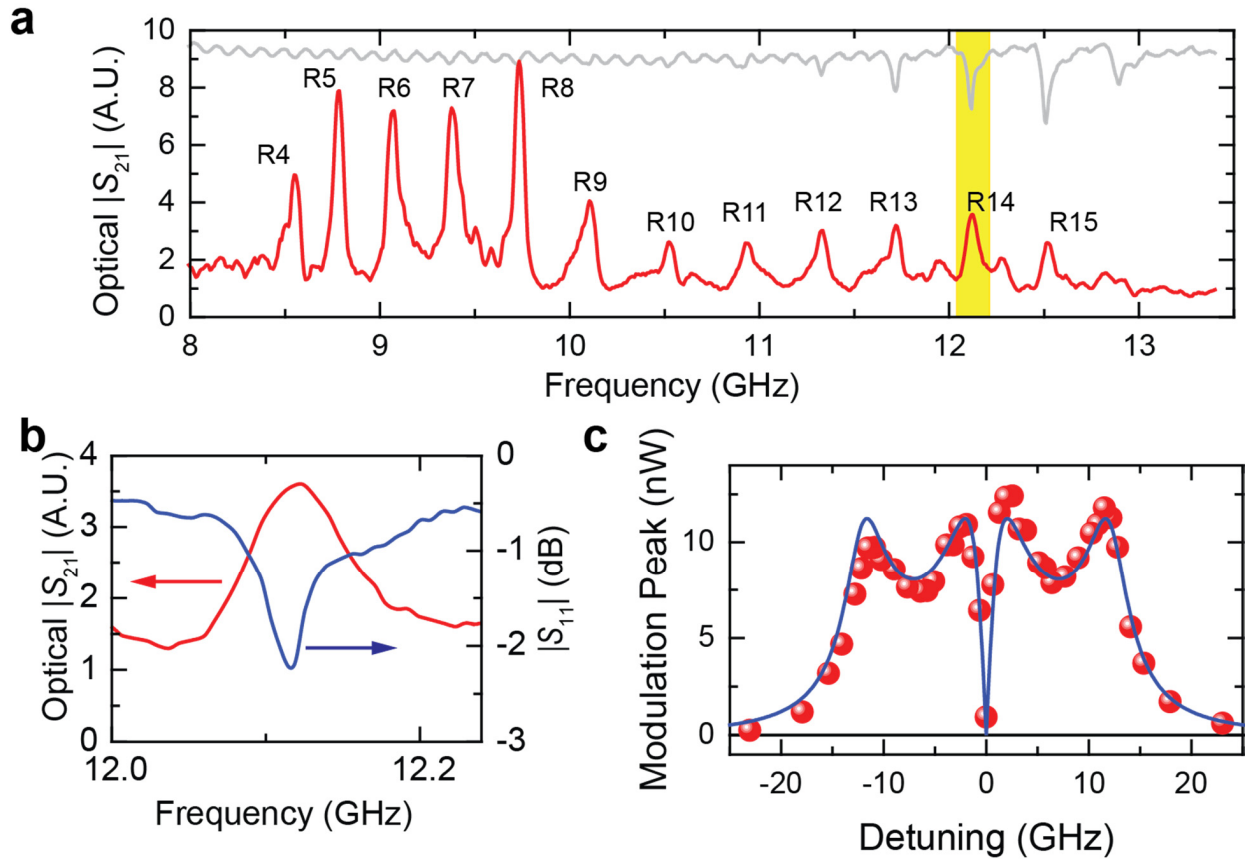

**Figure 2 Optomechanical modulation of the nanocavity by SAW wave in sideband resolve regime. a)** Spectrum of system transmission coefficient $S_{21}$ (red line) in linear scale, measured using optical detection with the nanocavity and electromechanical excitation of the SAW. Rayleigh modes (R4-R10) not visible in the reflection spectrum (grey line) can be detected with high sensitivity by the nanocavity. **b)** Zoom-in of the reflection and transmission spectra of R14 mode (inside the yellow box in panel **a**). **c)** Amplitude (peak value) of the oscillating optical power at the $S_{21}$ peak of R14 mode when laser detuning relative to the cavity resonance is varied. The data (red symbols) is fitted with theoretical model (blue line) of cavity optomechanical system in the sideband resolved regime (See Supplementary Info for details).

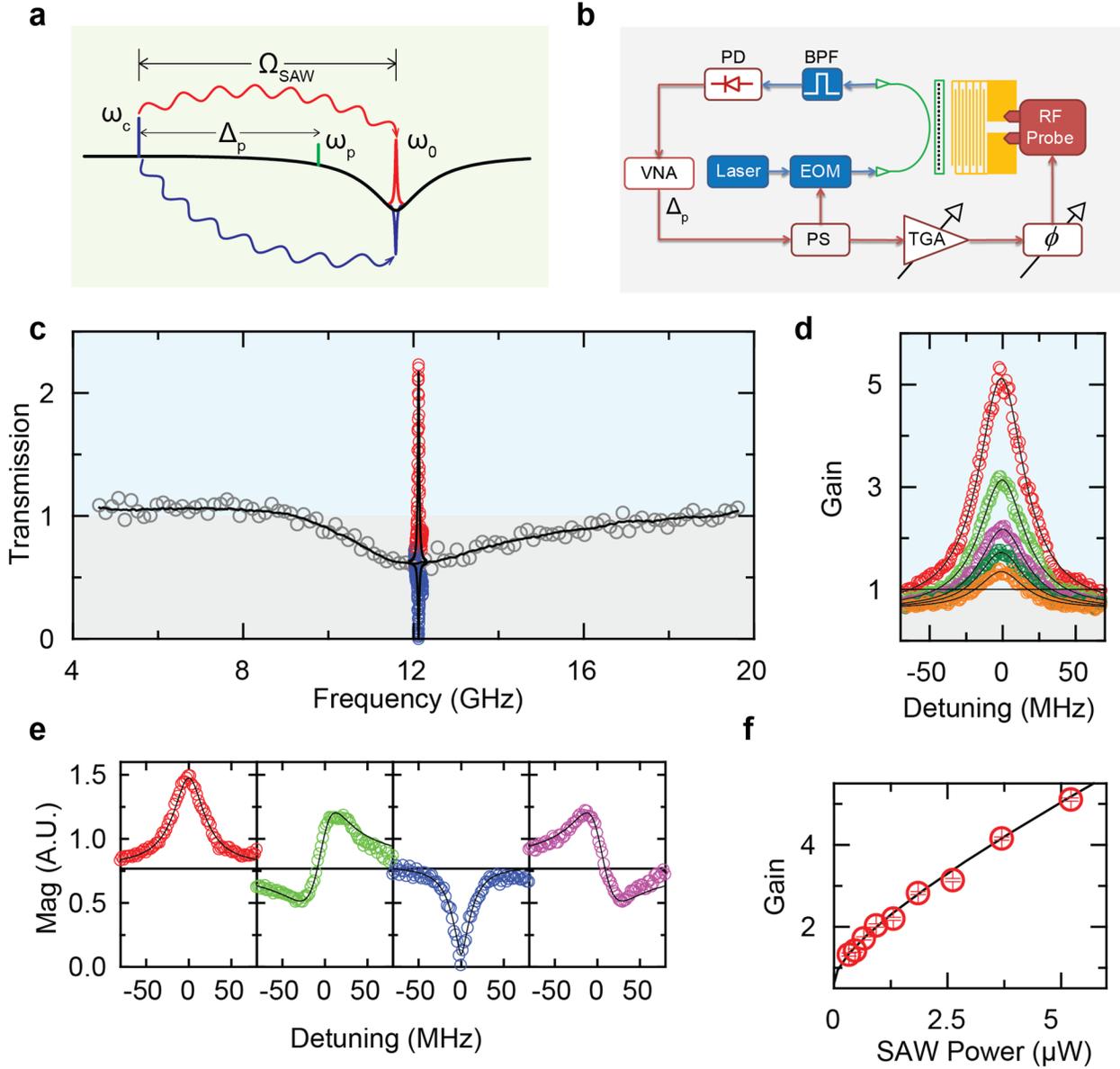

**Figure 3 Optomechanically induced transparency and absorption by propagating mechanical wave. a)** Diagram illustrating the three-wave mixing process of control ($\omega_c$), probe ($\omega_p$) and SAW waves ($\Omega_{SAW}$). The cavity resonance frequency is $\omega_0$ with a decay rate of $\kappa$. **b)** Homodyne measurement scheme used in the experiment. The probe light is derived from the control light when it is modulated at frequency $\Delta_p$, which is scanned to obtain the transmission spectrum. (NA: network analyzer; PS: power splitter; PD: photodetector; EOM: electro-optic modulator; TGA: tunable gain amplifier; $\phi$: phase shifter; BPF: band-pass filter.) **c)** Transmission spectrum of the probe light when the SAW is off (grey symbols) and on (red, blue symbols). Cavity

absorption is shown as the dip in the grey region. When the SAW wave induced anti-Stokes light is in-phase with the probe, constructive interference leads to transparency and gain as shown by the peak above unity transmission (the blue region). When the anti-Stokes light is $\pi$ out-of-phase with the probe, destructive interference enhances cavity absorption (the grey region), leading to high extinction of the probe. **d)** Gain of the system in the transparency region when the SAW power is increasing (orange: 0.33 µW, olive: 0.66 µW, purple: 1.3 µW, green: 2.6 µW, red: 5.2 µW). **e)** Transmitted probe light when the phase shift $\phi$ is set at 0 (red), $\pi/2$ (green), $\pi$ (blue), $3\pi/2$ (purple). When the phase is at $\pi/2$ and $3\pi/2$, the lineshapes imitate that of Fano resonances. **f)** The dependence of the system gain on the SAW power. The red symbols are experimental data while the black curve is the theoretical fitting.

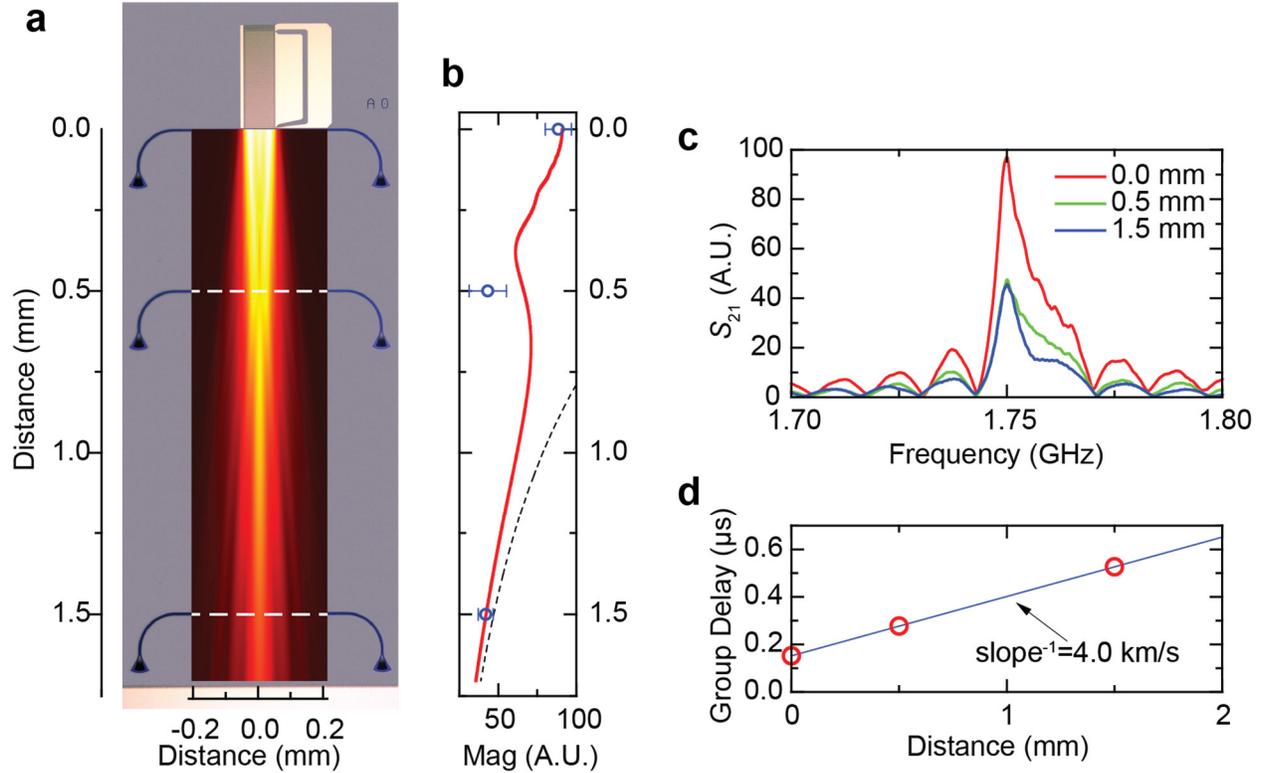

**Figure 4 SAW wave coupling with multiple cavities and its diffraction. a)** Optical image of a device with three nanocavities in the path of SAW wave propagation. The photonic cavities are end-coupled with the waveguides and the SAW operates at a lower frequency of 1.75 GHz. Overlaid on the image is the calculated amplitude distribution of the diffraction pattern of the SAW wave. **b)** The calculated displacement amplitude along the center line of the SAW beam, showing non-monotonic variation along the propagation direction. Symbols are $S_{21}$ magnitude measured from the three nanocavities. Dashed line is $e^{-\alpha r}/r^{1/2}$ asymptote of far-field amplitude of the wave for comparison, where $\alpha$ is the material loss assumed to be $(1.5\text{mm})^{-1}$. **c)** $S_{21}$ spectra measured from the three nanocavities at a distance of 0, 0.5 and 1.5 mm to the IDT. **d)** Group delay of the three cavities' responses to the SAW wave as a function of their distance to the IDT. The inverse of the slope gives group velocity of 4.0 km/s.

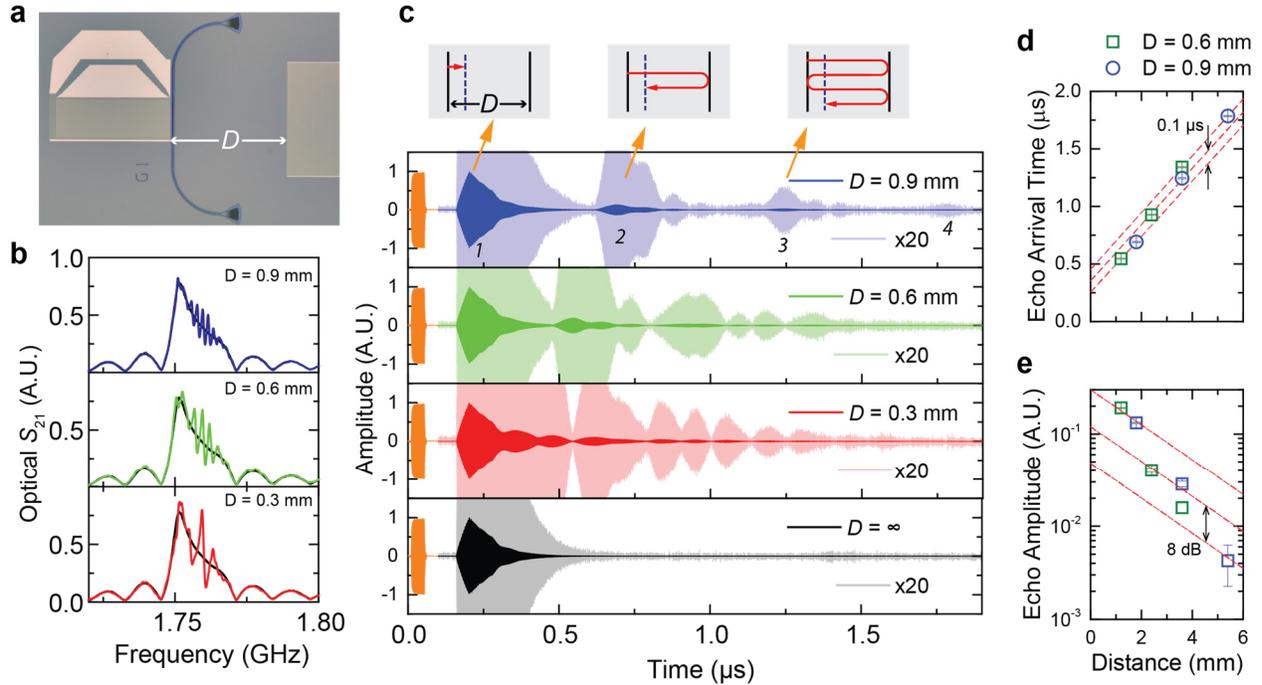

**Figure 5 A photonic cavity inside a phononic cavity or acoustic echo chamber. a)** Optical image of the device. The distance between the IDT and the Bragg reflector is $D$. **b)** $S_{21}$ spectra of devices with varying length $D$ of the phononic cavities (red, green blue lines), compared with a device without the Bragg reflector (black line). The spectra show peaks corresponding to the resonances of the phononic cavity with decreasing peak spacing (or free spectral range) when the cavity length is increased. **c)** Time-domain echo measurement of an acoustic pulse travelling inside phononic cavities of varying length $D$. Light colored traces are 20 times magnification of the dark colored traces. The acoustic pulse is excited by a 40 ns long burst of microwave at 1.75 GHz (orange). Up to four echoes of the pulse can be detected by the nanocavity. **d)** and **e)** The arrival time (**d**) and amplitude (**e**) of the detected echoes as a function of the apparent travel distance of the acoustic pulse. Red dashed lines are guides of the eyes assuming constant group velocity of 4.0 km/s (**d**) and exponential loss (**e**). The data deviate from the linear propagation due to extra delay (~0.1 μs) and loss (~8 dB) at the Bragg reflector of finite length.